# Series studies of the Potts model. II: Bulk series for the square lattice

K M Briggs†¶, I G Enting‡ and A J Guttmann†

†Department of Mathematics, The University of Melbourne, Parkville, Vic. Australia 3052.

‡CSIRO, Division of Atmospheric Research, Private Bag 1, Mordialloc, Vic. Australia 3195.

¶Present address: Department of Applied Mathematics, University of Adelaide, SA, Australia 5005.

Short title: *Square lattice Potts model*

**Abstract.** The finite lattice method of series expansion has been used to extend low-temperature series for the partition function, order parameter and susceptibility of the $q$-state Potts model to order $z^{56}$ (i.e. $u^{28}$), $z^{47}$, $z^{43}$, $z^{39}$, $z^{39}$, $z^{39}$, $z^{35}$, $z^{31}$ and $z^{31}$ for $q = 2, 3, 4, \ldots 9$ and 10 respectively. These series are used to test techniques designed to distinguish first-order transitions from continuous transitions. New numerical values are also obtained for the $q$-state Potts model with $q > 4$.

# 1. Introduction

This is the second in a series of papers in which we study the critical behaviour of the $q$-state Potts model in both two and three dimensions using series expansions derived from the finite lattice method. The previous paper (Guttmann and Enting, 1993), denoted I hereafter, gave the general expressions used to derive high- and low-temperature expansions for the $q$-state Potts model. In I, series expansions for the $q = 2$ (Ising) case on the simple cubic lattice were analysed. The present paper derives and analyses series for the bulk thermodynamic properties for Potts model on the square lattice for integer $q$ ranging from 2 to 10.

A brief history of the model follows: After the initial paper by Potts (1952), the model attracted little attention for almost two decades. During the 1970's there was greatly renewed interest, with new exact results, series studies and renormalisation group calculations as well as applications to phase transitions in surface films. A particular concern at that time was the failure of renormalisation group calculations to reproduce the exact results for the order of the transition in two dimensions. A review by Wu (1982) described much of the work on the Potts model.

The main exact results come from Potts (1952) and Baxter (1973, 1982). In particular, Potts (1952) located the critical temperature exactly for the two-dimensional Potts model on a square lattice by duality arguments. He found that $T_c = \Delta E/(k \ln(1 + \sqrt{q}))$ and $\bar{U} = \frac{1}{2}(U_c^+ + U_c^-) = \Delta E(1 - 1/\sqrt{q})$. (The energy is relative to the ground state energy, other aspects of the notation are defined in section 2 below).

For $q > 4$ the model has a first-order phase transition, and Baxter (1973) obtained the free energy and latent heat in 1973, and subsequently the value of the magnetisation at $T_c$ (Baxter, 1982). The results are listed in Tables 1 and 2 with the free energies converted to be consistent with our choice of zero ground-state energy. The free energy, $f$, defined by Baxter (1973) is related to the free energy, $F$, that we define below through the dimensionless form $F/\Delta E = f/\Delta E + 2$. Further, Baxter showed that the values of the magnetisation at $T_c$ were the same for the square, triangular and honeycomb lattices — a consequence of the star-triangle relation. For $q \le 4$ the model has a second-order phase transition. Indeed, for $q = 2$ the model is just the usual spin-$\frac{1}{2}$ Ising model, while $q = 4$ is the 'marginal' $q$ value, at which the exponents have confluent logarithmic corrections.

The critical exponents and critical temperature for $q \leq 4$ are shown in Table 3 below. The thermal exponent was given by Black and Emery (1981), following a conjecture of den Nijs (1979). Den Nijs (1983) also obtained the magnetic exponent, while Nienhuis (1982) obtained the (thermal) correction-to-scaling exponent $\Delta_1$. The results are:

$$2 - \alpha = \frac{2}{y_t} = \frac{2(2-r)}{3(1-r)}$$

$$1 + \frac{1}{\delta} = \frac{2}{y_h} = \frac{8(2-r)}{(3-r)(5-r)}$$

$$\Delta_1 = \frac{4r}{3(1-r)}$$

with

$$0 \leq r \equiv \frac{2}{\pi} \cos^{-1}(\sqrt{q}/2) \leq 1 \qquad \text{for } 0 \leq q \leq 4$$

However, for $q > 4$ certain properties still remain unknown. These include the value of the specific heat and of the isothermal susceptibility at the critical temperature. Various surface critical exponents and critical values are also unknown.

Of even greater interest is the behaviour of the three-dimensional Potts model. As noted above, for the $q = 2$ (Ising) case, the low-temperature series and some high-temperature series have recently been extended in I. For $q = 3$, the three-dimensional Potts model is of particular interest as it constrains the order of the deconfinement transition in quantum chromodynamics. The key question is whether the $q = 3$, $d = 3$ Potts model transition is first- or second-order. This is discussed further in the third of our series of papers.

The fact that the critical behaviour is known in the two-dimensional case makes it an ideal 'test-bed' for methods to distinguish first-order from second-order phase transitions. In this paper we have extended the low-temperature (and by duality, the zero-field high-temperature) series for the two-dimensional model for $q = 2, 3, 4, \ldots$, 9, 10. By the use of the finite-lattice method (see I and references therein), quite substantial series extensions have been made. By using differential approximants (Guttmann, 1989) to integrate the series, we have been able to clearly distinguish between first-order and second-order phase transitions.

The layout of the remainder of the paper is as follows: In the next section we briefly describe the finite lattice method and the nature of the results we have obtained therefrom. In section 3 we analyse the data. In section 4 we present a discussion of the results.

## 2. Series expansions from the finite lattice method

The definitions and notation follow the usage of I. The standard $q$-state Potts model is defined on a lattice with each site having a 'spin' variable that takes on $q$ possible values (denoted '0' to $q-1$). An energy $\Delta E > 0$ is associated with each pair of interacting sites that are in different spin states, and an energy of 0 applies to pairs of interacting sites in the same state. We consider only the square lattice, with each site interacting only with its 4 nearest neighbours. Each site not in state '0' has an additional field energy $H$.

The thermodynamic quantities can be derived from the partition function, $Z$. We choose the normalisation such that the state with all sites in state '0' has zero energy. This particular normalisation of the partition function is commonly denoted $\Lambda$.

We work in terms of the expansion variables $z = \exp(-\Delta E/kT)$, $\mu = \exp(-H/kT)$ and the high-temperature variable $v = (1-z)/(1+(q-1)z)$.

For the square lattice, the high-temperature expansion for the partition function takes the form (see I for the general case):

$$\Lambda = q^{-1}(1+(q-1)z)^2 \Phi(v) = q(1+(q-1)v)^{-2} \Phi(v) \tag{1}$$

with

$$\Phi(v) = 1 + (q-1)v^4 + \ldots \tag{2}$$

For the low temperature expansion, we use a modified field variable $x = 1 - \mu$ and truncate at order $x^2$ so that the partition function is expressed as

$$\Lambda = \Lambda_0 + x\Lambda_1 + x^2\Lambda_2 + \ldots \tag{3}$$

The zero-field partition function is expanded

$$\Lambda_0(z) = \sum_n \lambda_n z^n \tag{4}$$

On the square lattice, the duality relation takes the form

$$\Lambda_0(x) = \Phi(x) \qquad \text{(square lattice only)} \tag{5a}$$

so that the transition point (assuming it is unique) occurs (Potts, 1952) at

$$z_c = v_c = \frac{1}{1+\sqrt{q}} \qquad \text{(square lattice only)}. \tag{5b}$$

The dimensionless free energy is given by

$$\frac{F}{\Delta E} = -\frac{kT}{\Delta E} \ln \Lambda \tag{6}$$

so that for low temperatures, the internal energy is given by

$$U = \Delta E \, z \frac{d\Lambda_0}{dz} / \Lambda_0 \tag{7}$$

the order parameter by

$$M = 1 + \frac{q}{q-1} \frac{\Lambda_1}{\Lambda_0} = \sum_n m_n z^n \tag{8}$$

and the susceptibility by

$$\chi = 2\frac{\Lambda_2}{\Lambda_0} - \frac{\Lambda_1}{\Lambda_0} - \left(\frac{\Lambda_1}{\Lambda_0}\right)^2 = \sum_n c_n z^n. \tag{9}$$

Note that in I the expansions were expressed in terms of $u = z^2$ as only even powers of $z$ occur for $q = 2$. For $q \geq 3$ an additional 'transverse' susceptibility can be defined (Straley and Fisher, 1973) but is not considered here.

For $T \geq T_c$ the internal energy is given by

$$U = \Delta E \frac{\nu}{2} \frac{q-1}{q}(1-v) - \Delta E \frac{(1-v)(1+(q-1)v)}{q} \frac{d}{dv} \ln \Phi(v) \tag{10}$$

where $\nu$ is the lattice co-ordination number (4 in this case). Series expansions for the Potts model on the square lattice had been obtained previously by a number of workers. Kihara *et al* (1954) obtained the general-$q$ zero-field free energy series to order $z^{16}$ (or equivalently to $v^{16}$ by virtue of duality). Straley and Fisher (1973) obtained the general-$q$ general-field low-temperature series to $z^{13}$. Enting (1974) analysed the field grouping for $q = 3$ to order $\mu^9$ but the series were not published directly, but rather as 'coded' partial generating functions, based on the formalism of Sykes *et al* (1965).

The 3-state square lattice Potts model was the first application of the finite lattice method (de Neef, 1975; de Neef and Enting, 1977). This work obtained the high-temperature expansion to order 23 (working in powers of $\Delta E/kT$). The finite lattice method was used subsequently to obtain low-temperature expansions for $\Lambda$ and $M$ for $q = 3$ to order $z^{31}$ (Enting, 1980). These series were extended to $z^{35}$ (and new series for $\chi$ added) by Adler et al (1982). The algorithm used in the present work is essentially unchanged from that used in the 1980 and 1982 studies, although the program has been modified to work with general integer values of $q$. The increased number of terms that we have obtained reflects the increase in available computing capacity over the last decade rather than any major change to our technique. Recently, a preprint by Bhanot et al (1993) gives series for $q = 3$ and $q = 8$ on the square lattice (and some new $d = 3$ series). They used a transfer-matrix method related to our approach but with a more complicated (and apparently less efficient) choice of boundary conditions for their finite lattices. Their square lattice series extend earlier results but are slightly shorter than those presented here.

The basic formulation of the finite lattice method approximates the partition function per site, $Z$, as

$$Z = \lim_{|\Gamma| \to \infty} Z_\Gamma^{1/|\Gamma|} \approx \prod_{\alpha \in A} Z_\alpha^{W(\alpha)} \qquad (11a)$$

where $\Gamma$ denoted a graph (with $|\Gamma|$ sites) which is allowed to become arbitrarily large and $A$ is a set of finite lattices, $\alpha$, with $A$ closed under the operation of intersection of finite lattices. For the square lattice, this general relation has the specific form:

$$Z = \lim_{N \to \infty} Z_{NN}^{1/N^2} \cong \prod_{[q,r] \in A} Z_{qr}^{W(q,r)} \qquad (11b)$$

where $Z_{qr}$ is the partition function of a rectangle of dimensions $q \times r$ sites. For low-temperature expansions, the $Z_{qr}$ are to be evaluated with a surrounding layer of fully-ordered sites. The weights $W(q, r)$ depend on the set, $A$, over which the product is taken. In approximations (11a, 11b) an appropriate choice of weights will give $Z$ as a series correct up to, but not including, the order of the first connected graph that will not fit into any of the rectangles of set $A$ (Enting, 1978a).

For low-temperature Potts model series, the appropriate finite lattices are rectangles of $q \times r$ sites, surrounded by a boundary of sites fixed in state '0'. Inspection of the low-temperature expansion of the Potts model shows that the limiting graphs are trees that do not double back in any direction: all lines drawn perpendicular to bonds of the lattice intersect such trees at most once. Such a tree can span a rectangle of size $q \times r$ with $q + r - 1$ sites and $q + r - 2$ bonds in the tree and will give powers of $2(q+r)$ or more in the Potts model low-temperature variable, $z$. If one includes all rectangles such that $q + r \leq k$ then the series are correct to $z^{2k+1}$. We denote the set of rectangles with $q + r \leq k$ by $A(k)$.

The combinatorial factors from Enting (1978b) give

$$W(w, \ell) = \sum_{[q,r] \in A(k)} \eta(q - w)\, \eta(r - \ell) \qquad \text{for } [w, \ell] \in A(k) \tag{12}$$

where

$$\eta(0) = 1 \tag{13a}$$
$$\eta(1) = -2 \tag{13b}$$
$$\eta(2) = 1 \tag{13c}$$
$$\eta(k) = 0 \qquad \text{otherwise.} \tag{13d}$$

This implies

$$W(w, \ell) = 1 \qquad \text{for } w + \ell = k \tag{14a}$$
$$= -3 \qquad \text{for } w + \ell = k - 1 \tag{14b}$$
$$= 3 \qquad \text{for } w + \ell = k - 2 \tag{14c}$$
$$= -1 \qquad \text{for } w + \ell = k - 3 \tag{14d}$$
$$= 0 \qquad \text{otherwise} \tag{14e}$$

In actual computation it is convenient to exploit the symmetry and consider only $w \leq \ell$. We define $B(k) = \{[q, r] : q + r \leq k, q \leq r\}$. The expansion becomes

$$Z \approx \prod_{[q,r] \in B(k)} Z_{qr}^{V(q,r)} \tag{15}$$

with

$$V(\ell, w) = 2W(\ell, w) \qquad \text{for } w < \ell \tag{16a}$$
$$V(w, w) = W(w, w) \tag{16b}$$
$$V(\ell, w) = 0 \qquad \text{for } w > \ell \tag{16c}$$

Therefore we choose a maximum width $w_{\max}$ and work with $\ell + w \leq 2w_{\max} + 1$. This gives series correct to $z^{4w_{\max}+3}$.

The partition functions are constructed by using a transfer-matrix formalism to build up $\ell$ columns of length $w$. As in all of the most recent applications of the finite lattice method, we used the approach of building up the finite lattices one site at a time. The computational complexity of the calculation is determined by the largest value of $w$ that is required.

Storage is required for vectors giving the partial generating functions for all possible configurations of sites across a lattice. Without any simplification, such a vector will have $q^w$ elements for a rectangle of width $w$. Each element must have sufficient storage for a series truncated at the requisite order (i.e. $4(k+1)$ temperature terms times 3 field terms in the present case). Building up the lattice one site at a time means that the 'transfer matrix' is extremely sparse and the non-zero elements of the matrix can be calculated as required rather than having to be stored. The energies defined above only single out the '0' state and so the equivalence of the other states can be used to reduce the size of the vectors to approximately $q^w/(q-1)!$.

The precise size, $R(w, q)$ of vectors required to treat a lattice $w$ sites across is given in terms of $r_{wm}$ the number of ways of colouring $w$ sites with colours 0 to $q-1$, treating all permutations of colours 1 to $q-1$ as equivalent. That is

$$R(w, q) = \sum_{m=1}^{q} r_{wm} \qquad (17a)$$

with $r_{01} = r_{j1} = 1$ and the general relation

$$r_{jm} = r_{j-1,m} + m r_{j-1,m} \qquad (17b)$$

The series coefficients $\lambda_n$, $m_n$ and $c_n$ for $q = 2$ to $q = 10$ are listed in the tables in the appendix. For $q = 2$ the coefficients $\lambda_n$ and $m_n$ are known from the exact solutions. The series $c_n$ for $q = 2$ corrects two minor errors in the last two coefficients $c_{44}$ and $c_{46}$ obtained by Baxter and Enting (1979) using the corner-transfer-matrix technique. (The error in the work of Baxter and Enting arose from insufficient precision in the summation of their high-field polynomials — a full-precision summation of their published coefficients for the expansion (in powers of $u = z^2$ and $\mu$) gives the results obtained here).

Although the computer program used here is restricted to integer $q \geq 2$, the Potts model can be generalised to non-integer $q$ (Fortuin and Kasteleyn, 1972). A number of interesting special cases occur — in particular the limit $q \to 1$ (on any lattice) gives the statistics of the bond percolation problem (Fortuin and Kasteleyn, 1972; Wu, 1978). This connection has been exploited in series derivations (Enting, 1986). The finite lattice method is applicable to general $q$ and indeed one of the earliest applications of the method was in calculating the limit of chromatic polynomials which correspond to the $T \to 0$ limit of the antiferromagnetic Potts model, expressed as a function of $q$ (Kim and Enting, 1979).

## 3. Analysis of series

For a second-order phase transition, quantities such as the order parameter vanish at $T_c$, while quantities such as the susceptibility and specific heat diverge to infinity. For a first-order phase transition, all these quantities are expected to attain a finite, non-zero value at $T_c$, with finite slope at $T_c$. However, little is known rigorously about the nature of the transition. Among the possibilities are: (a) finite specific heats and susceptibilities allowing analytic continuation of the thermodynamic quantities beyond the transition point into a metastable region with a singularity $T_c^* > T_c$ on a 'pseudo-spinodal' line and effective 'critical exponents' at $T_c^*$; (b) finite specific heats and susceptibilities with a weak, essential singularity at $T_c$. Even with an essential singularity it may be possible to define the thermodynamic functions in the metastable region by analytic continuation in the complex plane passing *around* the singularity. (c) divergences in specific heats and susceptibilities (or their derivatives) at $T_c$. A previous attempt to use series expansions to search for an essential singularity gave inconclusive results (Enting and Baxter, 1980). Kim and Joseph (1975) presented evidence of 'possible diverging fluctuations at the first-order transition'. However none of our results seem to indicate any sort of singularity at the first-order transitions.

In analysing series expansions around the origin by Dlog Padé approximants or, more generally, differential approximants, poles and residues of the approximants will provide estimators of $T_c$ and the critical exponent in the case of a second-order phase transition, while in the case of a first-order transition, the approximant will furnish an effective analytic continuation, and provide estimators of $T_c^*$ and some effective exponent. If $T_c$ is exactly known, as it is for the two-dimensional Potts models, this observation provides an effective means to distinguish between the two types of phase transitions. We show this in Table 4, where we give the Dlog Padé approximants to the magnetisation series for the $q = 3$ Potts model, the $q = 5$ Potts model and the $q = 10$ Potts model. These are representative of a second-order transition, a weak first-order transition, and a first-order transition respectively. For the $q = 3$ case, we find $T_c^*/T_c = 0.99990$ and $\beta \approx 0.109$, compared to the exact result $\beta = 1/9$, so that the apparent critical temperature is less than 0.01% *below* the true critical temperature, while the critical exponent is correct to the quoted accuracy (the true value is 1/9 exactly). For the $q = 5$ Potts model, we find $T_c^*/T_c = 1.00064$ and $\beta \approx 0.077$, so that the apparent critical temperature is more than 0.06% *above* the true critical temperature, while the 'critical exponent' is rather erratically estimated as $\approx 0.077$ . For the $q = 10$ Potts model, we find $T_c^*/T_c = 1.0104$ and $\beta \approx 0.047$, so that the apparent critical temperature is more than 1.0% *above* the true critical temperature, while the 'critical exponent' is rather erratically estimated as $\approx 0.05$. (We emphasise that this so-called 'critical exponent' has no physical meaning).

This method of analysis alone appears to provide a reliable indicator of the order of a phase transition when the critical temperature is exactly known. For a second-order transition, the estimates of the critical temperature lie very slightly below $T_c$. (The $q = 2$, or Ising case for which the magnetisation can be represented exactly by low-order Dlog Padé approximants is an exception). Even for the marginal case of $q = 4$ (the critical dimension, where the model undergoes a second-order phase transition, but with logarithmic corrections to the critical exponents), we find $T_c^*/T_c = 0.99975$, and $\beta = 0.0906$, which is satisfyingly close to the exact value of $1/11$. For the weak first-order $q = 5$ case it is already clear that $T_c^*/T_c$ is significantly bigger than 1, while the 'exponent' estimates are much more erratic than for the $q \leq 4$ case.

The other numerical approach to distinguish between a first- and second-order phase transition is to compare numerical approximations to the free energy and internal energy in the high- and low-temperature regimes. (On the square lattice, we obtain the high-temperature series by duality from the low-temperature series). In Fig. 1 we show the plots of the free energy for $q = 3, 5$ and 10. For $q = 3$ the curve appears smooth, with no gradient discontinuities. At $q = 5$, a discontinuity in the gradient at $T_c$ is already apparent, while at $q = 10$ the discontinuity in the gradient is manifest. In Fig 2 we show the corresponding curves for the internal energy. The non-zero latent heat characteristic of a first-order transition is already manifest at $q = 5$. The numerical approximations used in these figures are the approximants formed by the method of differential approximants (DA) (Guttmann, 1989, page 83ff). This method generalizes Padé approximants by fitting an ordinary differential equation of the form
$$\sum_{i=0}^{m} Q_i(x) D^i f(x) = P(x)$$
(where $D^i = \frac{d^i}{dx^i}$) to the available series terms. Here $Q_k(x) = \sum_{i=0}^{m_k} q_{ki} x^i$ and $P(x) = \sum_{i=0}^{m_0} p_i x^i$ are polynomials. We chose $q_{m0} = 1$, so that the origin is not a regular singular point. This allows numerical integration of the differential equation starting at $x = 0$ in order to obtain the values plotted in the figures. For magnetisation series, homogeneous DAs ($P \equiv 0$) are often most useful. (For $m = 1$ this corresponds to logarithmic derivative Padé approximants). The degrees of $Q_k$ and $P$ are chosen to use all (or most) of the available series terms. In principle, any order of differential equations can be used, but first-order ($m = 1$) was mostly used in the current work. Finding the coefficients of $Q_k$ and $P$ reduces to solving a system of linear equations, but this system is often ill-conditioned, so that care must be taken in its solution.

This differential equation is then integrated numerically to obtain estimates of the desired physical quantities. In all cases a number (typically 10) of DAs using all the available coefficients were integrated. These were then averaged to obtain the means and standard deviations shown in the tables and graphs below. All calculations were performed in quadruple precision (approximately 34 decimal places), so that all series terms could be represented without loss of precision.

We performed the numerical integration with an extrapolation method of the Bulirsch-Stoer type, as described by Hairer (1987, Section II.9). The integrations were performed in terms of the series expansion variable but results are expressed in terms of $kT/\Delta E$.

The integrations described above clearly allow us to *qualitatively* distinguish between a first- and second-order phase transition. A much more stringent test is to *quantitatively* reproduce the magnetisation gap $\Delta M$ and the latent heat $\Delta U$ for $q \geq 5$. These have been calculated by Baxter (1973, 1982). These exact results are shown in table 2. The agreement between numerical and exact results is, as might be expected, best for $q$ close to 10, and worst near $q = 4$. Further details of the calculations are given in the subsections below.

We now discuss our numerical results in greater detail.

*3.1. Free energy and internal energy*

We integrated the equations defining approximants to the free energy and the internal energy series from $T = 0$ and $T = \infty$ to $T_c$ in order to determine the critical value of the free-energy and the latent heat (for $q > 4$) from the discontinuity at $T_c$. For low temperatures, $F$ was derived from approximants to $\ln \Lambda$ in powers of $z$ while for high temperatures we use approximants to $F/kT - 2\Delta E/kT$ expanded in powers of $v$. For the energy, high-temperature and low-temperature approximants were constructed from the respective expansions (in powers of $v$ and $z$) to $U/\Delta E - 2$. For $q \leq 4$, table 1 compares the exact values of $F/\Delta E$ to the series estimates. For $q \geq 5$ table 5 gives the series estimates for $F_c/\Delta E$ and $\Delta U/\Delta E$ which should be compared to the exact values in table 2. The critical value of the free-energy is obtained very accurately for all values of $q$. As expected from the known exact results (Baxter, 1973), $U$ was found to be continuous at $T_c$ for $q \leq 4$. From the tables, it can be seen that the exact results for the latent heat are reproduced to within a few percent, except at $q = 5$. However at $q = 5$ the latent heat is found to be larger than the exact value, which makes the order of the phase transition *more* obvious.

The behaviour of the free energy and internal energy as functions of temperature is shown in Figures 1 and 2 and 3. Since our primary objective is to test techniques that are applicable in three dimensions where $T_c$ is unknown, Fig. 3 which shows details of the transition region for $q = 5$ is of particular interest. The figure shows individual approximants, indicating the spread that occurs in the free energy itself and the precision with which the transition could be located if $T_c$ were not known.

*3.2. Magnetisation*

We integrated the approximants to the magnetisation series $M(z)$ from $T = 0$ to $T = T_c$ in order to determine the discontinuity $\Delta M$ at $T_c$. As expected from the known exact results (Baxter, 1982), $M$ vanished at $T_c$ for $q \leq 4$. Comparing tables 2 and 5, it can be seen that the exact results are reproduced to within a few percent, except at $q = 5$. At $q = 5$ the error in the magnetisation discontinuity is some 27%, falling to less than 1.5% at $q = 10$. A plot of the magnetisation for several values of $q$ is shown in Figure 4.

*3.3. Susceptibility and Specific Heat*

The susceptibility and specific heat properties are not known in general. Certainly for $q \leq 4$ they are known to diverge at $T_c$, but for $q \geq 5$ the behaviour is less well understood. Nevertheless, it is expected that the susceptibility and specific heat should remain finite at $T_c$, though this has not been proved. In order to study these rapidly increasing quantities, various sequence transformation were used to generate the most appropriate series, and hence DA, for numerical integration. In general, if a quantity $f(x)$ behaves at the origin of integration like $x^k$, it is usually desirable to remove this term and study $f(x)/x^k$ instead. In addition, if a function is increasing rapidly, but not necessarily diverging, studying the reciprocal of the function frequently provides better converged approximants. These two transformations are often used together so that, for example, we worked with the series for $z^4/\chi(z)$ (where $\chi$ is the susceptibility), rather than $\chi(z)$ itself.

In Table 6 we show the results for the specific heat and susceptibility at the critical temperature, when approached from both the high and low temperature side. We observe a monotonic decrease in the value of the specific heat at $T_c$ with increasing $q$. For $q = 5$ and $q = 6$ the integration is too unreliable to quote a result. The numerical evidence gives no suggestion of asymmetry in the specific heat values above and below $T_c$, though our error bars are too large to give a useful test of symmetry.

In connection with the error bars, the value for $C_0(T_c^-)$ for $q = 7$ seems anomalously low and yet, as for all the other cases, the range reflects the spread of approximants that were fitted. This anomaly serves to emphasise the fact that the ranges are obtained empirically, rather than being based on any statistical theory, and in addition reflect a relatively small number of cases.

Recently Billoire *et al* (1992) published a Monte Carlo study of the $q = 10$ Potts model specific heat. They obtained a value of 12.3 for $C_0(T_c)$ compared to our estimate of 32. However re-analysis (Billoire, private communication) indicates that the published value is a serious underestimate, due to the fact that his system was not large enough to eliminate finite size effects. However he is able to estimate the difference between the specific heats on the ordered and disordered side of the critical temperature and finds a value of 0.447 at $q = 10$ and 0.223 at $q = 7$. Such small differences, if they exist, cannot be resolved by our numerical estimates.

For the susceptibility we also observe the same monotonic trend of decreasing values at $T_c$ with increasing values of $q$. As far as we are aware, this is the first study of this quantity.

## 4. Discussion of results

We have shown how the finite lattice method may be used to extend Potts model series, and have used the method to substantially extend a number of series for a range of $q$ values. The series could all be extended by several further terms (typically 4) without excessive demand on computing resources, but we did not consider this necessary for our purposes. We have used the series, combined with appropriate numerical techniques, to show how a first-order phase transition can be distinguished from a second-order transition. In this way we find extremely strong evidence for the known first-order transition for $q \geq 5$. The methods developed in this paper are used in a subsequent paper to investigate the nature of the phase transition for the three-state 3-dimensional Potts model.

## Acknowledgments

Financial support from the Australian Research Council is acknowledged. IGE wishes to thank Doochul Kim for pointing out the inconsistency in the series published by Baxter and Enting. The authors wish to thank John O'Brien, Robert Bursill and Debbie Wood for assistance with some of the numerical work.

# Appendix 1. The series

Tables 7 to 15 list the series expansions that we have calculated.

**Figure captions**

**Figure 1.** Dimensionless free energy, $F/\Delta E$, of the square lattice Potts model for $q = 3$ (solid curve), $q = 5$ (short dashes) and $q = 10$ (long dashes) illustrating the change from continuous to first-order transitions. Curves are the average of typically 10 differential approximants.

**Figure 2.** Dimensionless internal energy, $U/\Delta E$, of the square lattice Potts model for $q = 3$ (solid curve), $q = 5$ (short dashes) and $q = 10$ (long dashes). Each curve is the average of typically 10 differential approximants.

**Figure 3.** Detail of the approximants to the Potts model free energy $(F/\Delta E)$ around the transition for $q = 5$. The plot includes 7 high-temperature approximants (shown as dashed) and 13 low-temperature approximants (shown as solid) although not all of these can be distinguished on this scale.

**Figure 4.** Spontaneous magnetisation of the square lattice Potts model for $q = 3, 5$ and 10. Each curve is the average of typically 10 differential approximants.

Tables and table captions

| $q$ | $z_c$ | $U_c/\Delta E$ | $F_c/\Delta E$ | $F_c/\Delta E$(series) |
|---|---|---|---|---|
| 2 | 0.414214 | 0.292893 | −0.054826 | −0.054825 |
| 3 | 0.366035 | 0.422650 | −0.059780 | −0.059777 |
| 4 | 0.333333 | 0.500000 | −0.056708 | −0.056722 |

Table 1. Exact critical properties for $q \leq 4$, from the work of Potts (1952) and Baxter (1973). The internal energies $U$ and free energy $F$ are defined relative to a ground state energy of zero. Numerical estimates of $F_c/\Delta E$ are also shown.

| $q$ | $kT_c/\Delta E$ | $\Delta M$ | $\Delta U/\Delta E$ | $\bar{U}/\Delta E$ | $F_c/\Delta E$ |
|---|---|---|---|---|---|
| 5 | .851528 | .492141 | .052919 | 0.552786 | −0.05205 |
| 6 | .807607 | .665181 | .201464 | 0.591752 | −0.04738 |
| 7 | .773059 | .749565 | .353277 | 0.622036 | −0.04311 |
| 8 | .744904 | .799837 | .486358 | 0.646447 | −0.03935 |
| 9 | .721348 | .833261 | .599668 | 0.666667 | −0.03608 |
| 10 | .701232 | .857107 | .696049 | 0.683772 | −0.03323 |

Table 2. Exact properties at the transition point for $q \geq 5$, from the work of Potts (1952) and Baxter (1973, 1982). The energies and free energy are defined relative to a ground state energy of zero.

| $q$ | $\alpha$ | $\beta$ | $\gamma$ | $\delta$ | $\nu$ | $\eta$ | $\Delta_1$ |
|---|---|---|---|---|---|---|---|
| 2 | 0 | 1/8 | 7/4 | 15 | 1 | 1/4 | 4/3 |
| 3 | 1/3 | 1/9 | 13/9 | 14 | 5/6 | 4/15 | 2/3 |
| 4 | 2/3 | 1/12 | 7/6 | 15 | 2/3 | 1/4 | 0 |

Table 3. Exact critical exponents for two-dimensional Potts models.

| N | [N−1/N] | | [N/N] | | [N+1/N] | |
|---|---|---|---|---|---|---|
| a) 3-state Potts model spontaneous magnetization. $z_c = 0.366025..$, $\beta = 0.1111...$ | | | | | | |
| 17 | 0.36595 | (0.1078) | 0.36615 | (0.1155) | 0.36591 | (0.1074)* |
| 18 | 0.36593 | (0.1076)* | 0.36583 | (0.1074)* | 0.36594 | (0.1078) |
| 19 | 0.36598 | (0.1084) | 0.36599 | (0.1086) | 0.36599 | (0.1086) |
| 20 | 0.36599 | (0.1086) | 0.36599 | (0.1086) | 0.36599 | (0.1086) |
| 21 | 0.36599 | (0.1086) | | | | |
| b) 5-state Potts model spontaneous magnetization. $z_c = 0.309016...$ | | | | | | |
| 15 | 0.30919 | (0.0768) | 0.30915 | (0.0763)* | 0.30917 | (0.0766)* |
| 16 | 0.30921 | (0.0772) | 0.30921 | (0.0772) | 0.30922 | (0.0772) |
| 17 | 0.30921 | (0.0772)* | 0.30921 | (0.0772)* | 0.30922 | (0.0772)* |
| 18 | 0.30921 | (0.0772)* | 0.30921 | (0.0772)* | 0.30920 | (0.0769)* |
| 19 | 0.30921 | (0.0772)* | | | | |
| c) 10-state Potts model spontaneous magnetization. $z_c = 0.240253...$ | | | | | | |
| 11 | 0.24280 | (0.0436) | 0.24124 | (0.0346)* | 0.24288 | (0.0441)* |
| 12 | 0.24288 | (0.0441)* | 0.24278 | (0.0433)* | 0.24269 | (0.0428)* |
| 13 | 0.24269 | (0.0428)* | 0.24271 | (0.0430)* | 0.24267 | (0.0427)* |
| 14 | 0.24268 | (0.0427)* | 0.24275 | (0.0432)* | 0.24276 | (0.0433)* |
| 15 | 0.24276 | (0.0433)* | | | | |

Table 4. Singularities estimated from Padé approximants to spontaneous magnetisation of the Potts model for $q = 3, 5$ and $10$

| q | $kT_c/\Delta E$ | $\Delta M$ | $\Delta U/\Delta E$ | $F_c/\Delta E$ |
|---|---|---|---|---|
| 5 | .85152841 | .643 ± 0.002 | 0.085 ± 0.015 | −0.05204 |
| 6 | .80760682 | .728 ± 0.010 | 0.200 ± 0.010 | −0.04758 |
| 7 | .77305889 | .775 ± 0.010 | 0.375 ± 0.027 | −0.04309 |
| 8 | .74490446 | .817 ± 0.003 | 0.499 ± 0.007 | −0.03951 |
| 9 | .72134751 | .846 ± 0.002 | 0.592 ± 0.014 | −0.03598 |
| 10 | .70123160 | .862 ± 0.030 | 0.706 ± 0.005 | −0.03320 |

Table 5. Square lattice Potts model, $\Delta M$, $\Delta U$ and $F_c$ results from series analysis. Compare to exact values in Table 2. Free energies are defined relative to a ground-state energy of zero.

| q | $kT_c/\Delta E$ | $\chi(T_c^-)$ | $C_0(T_c^-)$ | $C_0(T_c^+)$ |
|---|---|---|---|---|
| 5 | .85152841 | 100 ± 50 | — | — |
| 6 | .80760682 | 36 ± 10 | — | — |
| 7 | .77305889 | 15 ± 4 | 93 ± 3 | 250 ± 160 |
| 8 | .74490446 | 7.3 ± 0.6 | 72 ± 20 | 75 ± 10 |
| 9 | .72134751 | 3.4 ± 0.9 | 40 ± 5 | 42 ± 4 |
| 10 | .70123160 | 2.44 ± 0.09 | 31.8 ± 2.8 | 33 ± 3 |

**Table 6.** Square lattice Potts model, susceptibility and specific heat results from series analysis

| | Partition function | Magnetisation | Susceptibility |
|---|---|---|---|
| 0 | 1 | 1 | 0 |
| 2 | 0 | 0 | 0 |
| 4 | 1 | -2 | 1 |
| 6 | 2 | -8 | 8 |
| 8 | 5 | -34 | 60 |
| 10 | 14 | -152 | 416 |
| 12 | 44 | -714 | 2791 |
| 14 | 152 | -3472 | 18296 |
| 16 | 566 | -17318 | 118016 |
| 18 | 2234 | -88048 | 752008 |
| 20 | 9228 | -454378 | 4746341 |
| 22 | 39520 | -2373048 | 29727472 |
| 24 | 174271 | -12515634 | 185016612 |
| 26 | 787246 | -66551016 | 1145415208 |
| 28 | 3628992 | -356345666 | 7059265827 |
| 30 | 17019374 | -1919453984 | 43338407712 |
| 32 | 81011889 | -10392792766 | 265168691392 |
| 34 | 390633382 | -56527200992 | 1617656173824 |
| 36 | 1905134695 | -308691183938 | 9842665771649 |
| 38 | 9385453576 | -1691769619240 | 59748291677832 |
| 40 | 46653815395 | -9301374102034 | 361933688520940 |
| 42 | 233788460256 | -51286672777080 | 2188328005246304 |
| 44 | 1180111379105 | -283527726282794 | 13208464812265559 |
| 46 | 5996452414310 | -1571151822119216 | 79600379336505560 |
| 48 | 30653752894948 | -8725364469143718 | 479025509574159232 |
| 50 | 157568531636534 | -48552769461088336 | 2878946431929191656 |
| 52 | 814062277383328 | -270670485377401738 | 17281629934637476365 |
| 54 | 4225485275503702 | -1511484024051198680 | 103621922312364296112 |
| 56 | 22027957435784967 | -8453722260102884930 | 620682823263814178484 |

**Table 7.** $q = 2$ square lattice Potts model

|    | Partition function      | Magnetisation         | Susceptibility         |
|----|-------------------------|-----------------------|------------------------|
| 0  | 1                       | 1                     | 0                      |
| 1  | 0                       | 0                     | 0                      |
| 2  | 0                       | 0                     | 0                      |
| 3  | 0                       | 0                     | 0                      |
| 4  | 2                       | -3                    | 2                      |
| 5  | 0                       | 0                     | 0                      |
| 6  | 4                       | -12                   | 16                     |
| 7  | 4                       | -12                   | 16                     |
| 8  | 6                       | -36                   | 100                    |
| 9  | 24                      | -108                  | 216                    |
| 10 | 24                      | -210                  | 844                    |
| 11 | 68                      | -480                  | 1552                   |
| 12 | 190                     | -1746                 | 7844                   |
| 13 | 192                     | -2340                 | 12112                  |
| 14 | 904                     | -10566                | 60268                  |
| 15 | 1420                    | -19500                | 118944                 |
| 16 | 3106                    | -53976                | 424072                 |
| 17 | 9940                    | -152604               | 1081392                |
| 18 | 14572                   | -329424               | 3201728                |
| 19 | 49268                   | -971304               | 8670688                |
| 20 | 102886                  | -2403291              | 25713154               |
| 21 | 225004                  | -5955576              | 67206560               |
| 22 | 652940                  | -16858584             | 203077760              |
| 23 | 1301256                 | -40337376             | 532881432              |
| 24 | 3513806                 | -110301321            | 1558159918             |
| 25 | 8591792                 | -287061696            | 4250639632             |
| 26 | 19326248                | -730223208            | 11956293152            |
| 27 | 52781148                | -1985703720           | 33296697848            |
| 28 | 120709472               | -5070001716           | 92820406096            |
| 29 | 306339824               | -13446444720          | 257249275776           |
| 30 | 779682608               | -35650214232          | 721023458656           |
| 31 | 1852672272              | -92442918828          | 1986080278600          |
| 32 | 4847112666              | -247542929499         | 5561045323298          |
| 33 | 11876028924             | -648347258796         | 15359165767512         |
| 34 | 29820747120             | -1713912378552        | 42717426328784         |
| 35 | 76592341404             | -4559593914288        | 118457421095792        |
| 36 | 189184240720            | -11991311519034       | 328170466563836        |
| 37 | 486960149980            | -31943103715128       | 909829346983664        |
| 38 | 1230269248240           | -84599939924118       | 2520622606225868       |
| 39 | 3111387440800           | -224265087762144      | 6973368153491880       |
| 40 | 8008990142050           | -597511883594619      | 19322697243220158      |
| 41 | 20253094484576          | -1584231404110704     | 53409977638363032      |
| 42 | 52022867385004          | -4220295103426356     | 147820297067842856     |
| 43 | 133290716187904         | -11234571367790256    | 408655295665071080     |
| 44 | 340509251651724         | -29892611571334848    | 1129521213462962520    |
| 45 | 878668731837260         | -79763126301078204    | 3122011116123891464    |
| 46 | 2252826675055124        | -212500082474434470   | 8624059746484047468    |
| 47 | 5806881993986032        | -567062477783225940   | 23820051913808354000   |

**Table 8.** $q = 3$ square lattice Potts model

|    | Partition function     | Magnetisation        | Susceptibility      |
|----|------------------------|----------------------|---------------------|
| 0  | 1                      | 1                    | 0                   |
| 1  | 0                      | 0                    | 0                   |
| 2  | 0                      | 0                    | 0                   |
| 3  | 0                      | 0                    | 0                   |
| 4  | 3                      | -4                   | 3                   |
| 5  | 0                      | 0                    | 0                   |
| 6  | 6                      | -16                  | 24                  |
| 7  | 12                     | -32                  | 48                  |
| 8  | 3                      | -28                  | 120                 |
| 9  | 72                     | -288                 | 648                 |
| 10 | 66                     | -400                 | 1608                |
| 11 | 144                    | -1024                | 4176                |
| 12 | 822                    | -5268                | 21093               |
| 13 | 480                    | -5920                | 38064               |
| 14 | 3624                   | -32160               | 175608              |
| 15 | 8508                   | -82720               | 494616              |
| 16 | 10482                  | -163020              | 1365726             |
| 17 | 65856                  | -737568              | 5077200             |
| 18 | 94794                  | -1482784             | 13549704            |
| 19 | 289452                 | -4644992             | 43359768            |
| 20 | 1008420                | -15095436            | 140590629           |
| 21 | 1561032                | -33307648            | 389348688           |
| 22 | 6503532                | -117747376           | 1296882504          |
| 23 | 15224016               | -312435552           | 3834279072          |
| 24 | 34976979               | -842726356           | 11499126642         |
| 25 | 125988144              | -2747491616          | 36680416368         |
| 26 | 263308986              | -7020371952          | 107193301920        |
| 27 | 805096764              | -21348043296         | 333178056720        |
| 28 | 2319752694             | -62732977996         | 1019415082779       |
| 29 | 5402283396             | -169814283264        | 3037827148632       |
| 30 | 17415097542            | -524175339168        | 9438120599520       |
| 31 | 44310604860            | -1465377774880       | 28340399493144      |
| 32 | 120262240257           | -4227843277380       | 86034549347280      |
| 33 | 361259402196           | -12642298828704      | 263586587279472     |
| 34 | 915351056190           | -35439363555136      | 791349060376776     |
| 35 | 2690038490904          | -105238706111616     | 2417035981737624    |
| 36 | 7502832907557          | -305746580682940     | 7324176466760445    |
| 37 | 20120170776144         | -877576741412064     | 22116375075991056   |
| 38 | 59297057120916         | -2604469134327440    | 67378910515598736   |
| 39 | 160763118088260        | -7498478213381792    | 203361542589030720  |
| 40 | 453982230850713        | -21932701743507244   | 616448061791922708  |
| 41 | 1302433413699684       | -64438624831555744   | 1869466917535240656 |
| 42 | 3570759048806208       | -186505497431699280  | 5644507206242675400 |
| 43 | 10294158055979004      | -549250323031345792  | 17116665818567515608|

**Table 9.** $q = 4$ square lattice Potts model

|    | Partition function | Magnetisation | Susceptibility |
|----|---|---|---|
| 0  | 1 | 1 | 0 |
| 1  | 0 | 0 | 0 |
| 2  | 0 | 0 | 0 |
| 3  | 0 | 0 | 0 |
| 4  | 4 | -5 | 4 |
| 5  | 0 | 0 | 0 |
| 6  | 8 | -20 | 32 |
| 7  | 24 | -60 | 96 |
| 8  | -4 | -10 | 120 |
| 9  | 144 | -540 | 1296 |
| 10 | 176 | -830 | 3032 |
| 11 | 168 | -1440 | 7392 |
| 12 | 2348 | -12930 | 48856 |
| 13 | 1200 | -12660 | 86496 |
| 14 | 8792 | -72250 | 405368 |
| 15 | 34056 | -266220 | 1473600 |
| 16 | 21092 | -364490 | 3423800 |
| 17 | 249768 | -2407020 | 16113120 |
| 18 | 466952 | -5493880 | 45751360 |
| 19 | 894840 | -14148000 | 141321696 |
| 20 | 5545356 | -66328975 | 563442380 |
| 21 | 7573416 | -133669680 | 1532848896 |
| 22 | 31825552 | -507900420 | 5523490864 |
| 23 | 109857648 | -1701343560 | 18731006352 |
| 24 | 183834532 | -4003389435 | 55057588668 |
| 25 | 911149824 | -16157898840 | 201014668032 |
| 26 | 2193242320 | -44696695560 | 626745614848 |
| 27 | 5622993528 | -131231068680 | 2011978174032 |
| 28 | 22900219536 | -476230251170 | 6990596535720 |
| 29 | 49840002048 | -1272406453680 | 21530011859136 |
| 30 | 170996310488 | -4263523739780 | 72383200786800 |
| 31 | 547847760000 | -13708308759180 | 238951933550064 |
| 32 | 1328084520588 | -38993908456195 | 754059402907804 |
| 33 | 4859021632872 | -133683456488820 | 2546174053634736 |
| 34 | 13419413642968 | -401454864043460 | 8178788230326672 |
| 35 | 38470066484088 | -1235148320717160 | 26521657626107232 |
| 36 | 131314151349976 | -4088698146399590 | 88159797282578792 |
| 37 | 351804517490808 | -12164467726392600 | 282022546158446016 |
| 38 | 1124247366814936 | -39182468233883110 | 927533183411394600 |
| 39 | 3501627546572496 | -124429743346539840 | 3032466102194828496 |

**Table 10.** $q = 5$ square lattice Potts model

|    | Partition function | Magnetisation | Susceptibility |
|----|---|---|---|
| 0  | 1 | 1 | 0 |
| 1  | 0 | 0 | 0 |
| 2  | 0 | 0 | 0 |
| 3  | 0 | 0 | 0 |
| 4  | 5 | -6 | 5 |
| 5  | 0 | 0 | 0 |
| 6  | 10 | -24 | 40 |
| 7  | 40 | -96 | 160 |
| 8  | -15 | 18 | 100 |
| 9  | 240 | -864 | 2160 |
| 10 | 390 | -1608 | 5440 |
| 11 | 80 | -1536 | 10720 |
| 12 | 5200 | -26478 | 97835 |
| 13 | 3120 | -26208 | 172960 |
| 14 | 16000 | -131472 | 781960 |
| 15 | 100360 | -695136 | 3635280 |
| 16 | 35830 | -705282 | 7301860 |
| 17 | 676000 | -6069600 | 40513440 |
| 18 | 1760530 | -17145840 | 130258520 |
| 19 | 1919240 | -33196224 | 362286640 |
| 20 | 20588740 | -219104358 | 1769583805 |
| 21 | 31455520 | -454883520 | 4908084320 |
| 22 | 102369320 | -1589271912 | 17743365920 |
| 23 | 534753600 | -7015071072 | 71015805120 |
| 24 | 750136775 | -14801255718 | 202223457160 |
| 25 | 4129522880 | -66670704288 | 808396021120 |
| 26 | 13033586990 | -217598415192 | 2788938475160 |
| 27 | 25901162520 | -571905903072 | 8810020871360 |
| 28 | 139208616380 | -2509604877582 | 34575265840675 |
| 29 | 331986591080 | -7057292732736 | 111570542543280 |
| 30 | 1007597125790 | -23164508689248 | 386492454099760 |
| 31 | 4245621181560 | -88855085448096 | 1424843663011600 |
| 32 | 9716604610065 | -248631471063042 | 4599792447092840 |
| 33 | 37562154772200 | -915400373994912 | 16638531775796160 |
| 34 | 125381210315310 | -3099289015999584 | 58025774058205040 |
| 35 | 327356630630880 | -9366267243722112 | 193486559359486160 |
| 36 | 1307950724960515 | -34832974106130846 | 700334584661741705 |
| 37 | 3808355804296960 | -110265822998373408 | 2373578442196407200 |
| 38 | 11776254793724920 | -362663928629146008 | 8176367783196681800 |
| 39 | 43485616464063400 | -1292915644379929056 | 29029320310487709600 |

**Table 11.** $q = 6$ square lattice Potts model

|    | Partition function | Magnetisation | Susceptibility |
|----|-------------------|---------------|----------------|
| 0  | 1 | 1 | 0 |
| 1  | 0 | 0 | 0 |
| 2  | 0 | 0 | 0 |
| 3  | 0 | 0 | 0 |
| 4  | 6 | -7 | 6 |
| 5  | 0 | 0 | 0 |
| 6  | 12 | -28 | 48 |
| 7  | 60 | -140 | 240 |
| 8  | -30 | 56 | 60 |
| 9  | 360 | -1260 | 3240 |
| 10 | 744 | -2842 | 9156 |
| 11 | -180 | -1120 | 13680 |
| 12 | 9834 | -47754 | 175116 |
| 13 | 7440 | -51940 | 319920 |
| 14 | 23952 | -207102 | 1329156 |
| 15 | 240180 | -1547980 | 7815840 |
| 16 | 65046 | -1295028 | 14115696 |
| 17 | 1478940 | -12844860 | 86758800 |
| 18 | 5278404 | -45718288 | 323797728 |
| 19 | 3298380 | -66714200 | 798016320 |
| 20 | 59010978 | -588028203 | 4634018046 |
| 21 | 110649780 | -1360197160 | 13634236320 |
| 22 | 253716060 | -4028581648 | 47009324592 |
| 23 | 1946334840 | -23146611600 | 221905374120 |
| 24 | 2709355266 | -47264561869 | 625484761002 |
| 25 | 13823742000 | -214532828720 | 2603227959120 |
| 26 | 58205917656 | -851529468216 | 10144975819968 |
| 27 | 96209517780 | -2020426505160 | 31145410317960 |
| 28 | 603180202992 | -10061684149216 | 134214946004952 |
| 29 | 1719615462960 | -31538686100880 | 464836718946240 |
| 30 | 4359818416824 | -96228336284304 | 1603857725501712 |
| 31 | 23080980564720 | -433388885935180 | 6571731067255800 |
| 32 | 55306170014214 | -1246332722187651 | 21917612779277742 |
| 33 | 201109899966180 | -4603433032789500 | 82056490326269640 |
| 34 | 826406173630152 | -17921451461635792 | 313584713235621024 |
| 35 | 2050391032558740 | -53354558614082880 | 1065067369034097360 |

**Table 12.** $q = 7$ square lattice Potts model

|    | Partition function | Magnetisation | Susceptibility |
|----|-------------------:|--------------:|---------------:|
| 0  | 1 | 1 | 0 |
| 1  | 0 | 0 | 0 |
| 2  | 0 | 0 | 0 |
| 3  | 0 | 0 | 0 |
| 4  | 7 | -8 | 7 |
| 5  | 0 | 0 | 0 |
| 6  | 14 | -32 | 56 |
| 7  | 84 | -192 | 336 |
| 8  | -49 | 104 | 0 |
| 9  | 504 | -1728 | 4536 |
| 10 | 1274 | -4640 | 14504 |
| 11 | -672 | 0 | 15792 |
| 12 | 16730 | -78696 | 288169 |
| 13 | 15792 | -96960 | 556752 |
| 14 | 30464 | -293056 | 2062088 |
| 15 | 497700 | -3061440 | 15132264 |
| 16 | 135506 | -2341976 | 25582802 |
| 17 | 2802240 | -24026304 | 165495792 |
| 18 | 13293602 | -106819264 | 720185368 |
| 19 | 5007828 | -122471424 | 1588846728 |
| 20 | 141712956 | -1357292536 | 10588862669 |
| 21 | 330138984 | -3611133696 | 33856668720 |
| 22 | 530780740 | -8864133984 | 108773186200 |
| 23 | 5749786896 | -64316947776 | 596266427232 |
| 24 | 8786258971 | -135447363816 | 1709093729238 |
| 25 | 37764443472 | -579146076096 | 7126592218032 |
| 26 | 207473441098 | -2784296946528 | 31462130162512 |
| 27 | 316373962548 | -6222589183680 | 94727643465168 |
| 28 | 2063246723202 | -32959913315576 | 433862737592571 |
| 29 | 7171329549804 | -118042425288960 | 1637231047784136 |
| 30 | 15593869260998 | -333304465512512 | 5555976600192816 |
| 31 | 96599146653492 | -1696556998769088 | 24859503635800680 |
| 32 | 258118001999361 | -5212673379084904 | 86979393078997048 |
| 33 | 842597112773724 | -18588152419905216 | 328938272280954672 |
| 34 | 4144519614837910 | -82387658612630912 | 1373037610865700024 |
| 35 | 10425182606664504 | -248254672744051968 | 4775687621042306472 |

**Table 13.** $q = 8$ square lattice Potts model

|    | Partition function | Magnetisation    | Susceptibility    |
|----|--------------------|------------------|-------------------|
| 0  | 1                  | 1                | 0                 |
| 1  | 0                  | 0                | 0                 |
| 2  | 0                  | 0                | 0                 |
| 3  | 0                  | 0                | 0                 |
| 4  | 8                  | -9               | 8                 |
| 5  | 0                  | 0                | 0                 |
| 6  | 16                 | -36              | 64                |
| 7  | 112                | -252             | 448               |
| 8  | -72                | 162              | -80               |
| 9  | 672                | -2268            | 6048              |
| 10 | 2016               | -7110            | 21808             |
| 11 | -1456              | 2016             | 16576             |
| 12 | 26392              | -121338          | 444848            |
| 13 | 30240              | -170100          | 919744            |
| 14 | 32464              | -379890          | 2987248           |
| 15 | 929488             | -5535180         | 27006336          |
| 16 | 297352             | -4188870         | 44211856          |
| 17 | 4789456            | -41053068        | 289598400         |
| 18 | 29374864           | -224389944       | 1462179584        |
| 19 | 7376432            | -213735312       | 2947217728        |
| 20 | 299855320          | -2801005551      | 21805200184       |
| 21 | 855405712          | -8621268768      | 76569774848       |
| 22 | 993563072          | -17694788796     | 228101077664      |
| 23 | 14579082336        | -156613057272    | 1422679507872     |
| 24 | 25470896744        | -353319727311    | 4234628561272     |
| 25 | 89577509504        | -1377018275976   | 17279449575808    |
| 26 | 621970314656       | -7875911823192   | 85777377714560    |
| 27 | 950903499504       | -17325165946536  | 257362505607200   |
| 28 | 5956796021984      | -92844355601646  | 1218813810751024  |
| 29 | 24987280793216     | -380787988636944 | 5033799620045952  |
| 30 | 49220523168272     | -1014469914963708| 16823743332996640 |
| 31 | 333054308294400    | -5604585935358060| 80496531779269984 |

**Table 14.** $q = 9$ square lattice Potts model

|    | Partition function | Magnetisation      | Susceptibility      |
|----|--------------------|--------------------|---------------------|
| 0  | 1                  | 1                  | 0                   |
| 1  | 0                  | 0                  | 0                   |
| 2  | 0                  | 0                  | 0                   |
| 3  | 0                  | 0                  | 0                   |
| 4  | 9                  | -10                | 9                   |
| 5  | 0                  | 0                  | 0                   |
| 6  | 18                 | -40                | 72                  |
| 7  | 144                | -320               | 576                 |
| 8  | -99                | 230                | -180                |
| 9  | 864                | -2880              | 7776                |
| 10 | 3006               | -1036              | 31392               |
| 11 | -2592              | 5120               | 15552               |
| 12 | 39348              | -177810            | 653391              |
| 13 | 53280              | -281920            | 1452096             |
| 14 | 25992              | -454800            | 4102488             |
| 15 | 1605456            | -9336640           | 45188640            |
| 16 | 631062             | -7348830           | 73492200            |
| 17 | 7576128            | -65490240          | 474281280           |
| 18 | 58741002           | -432882160         | 2754369000          |
| 19 | 11555280           | -362940800         | 5188189536          |
| 20 | 577400796          | -5306650050        | 41404785525         |
| 21 | 1976057856         | -18788133760       | 159998233536        |
| 22 | 1734448752         | -32994025240       | 444259126224        |
| 23 | 32942677248        | -344082515520      | 3087795261312       |
| 24 | 66351732687        | -846585038170      | 9670634655948       |
| 25 | 192109371264       | -2979703762880     | 38172808778112      |
| 26 | 1631012967630      | -19830863413320    | 210547287205128     |
| 27 | 2631716943408      | -44403893943360    | 639171550340352     |
| 28 | 15171912275256     | -233105484042490   | 3068237140062435    |
| 29 | 75277805632848     | -1085356198700160  | 13833211331827296   |
| 30 | 141748393419918    | -2799174780003360  | 45883584616400640   |
| 31 | 991716118694640    | -16218543098061760 | 2303350820283032264 |

**Table 15.** $q = 10$ square lattice Potts model

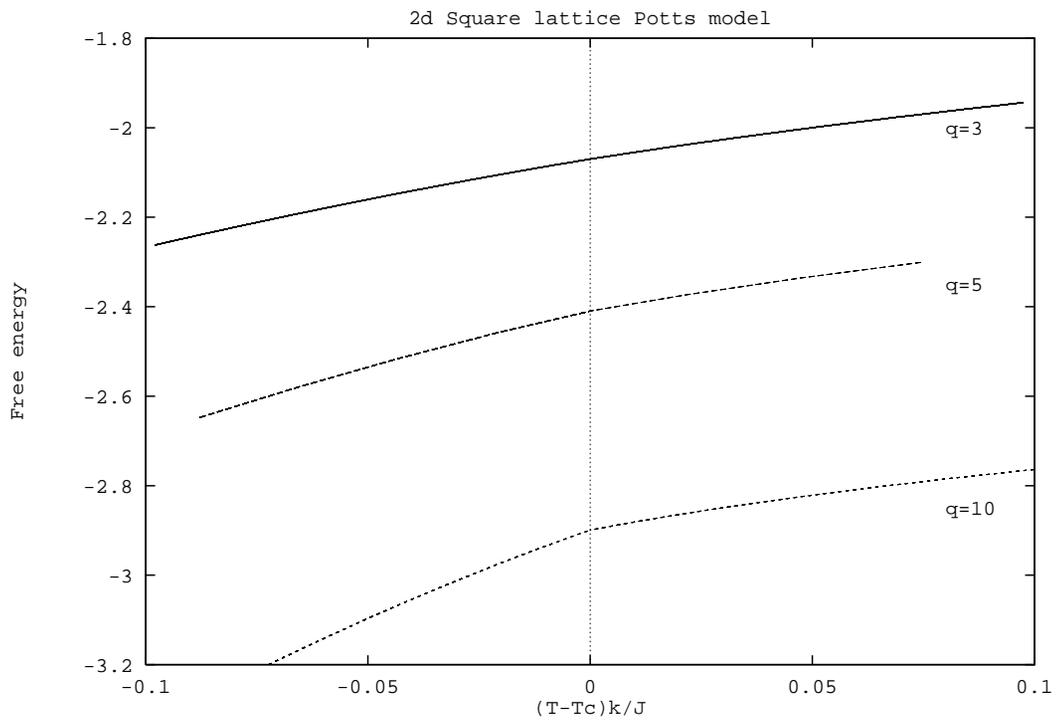

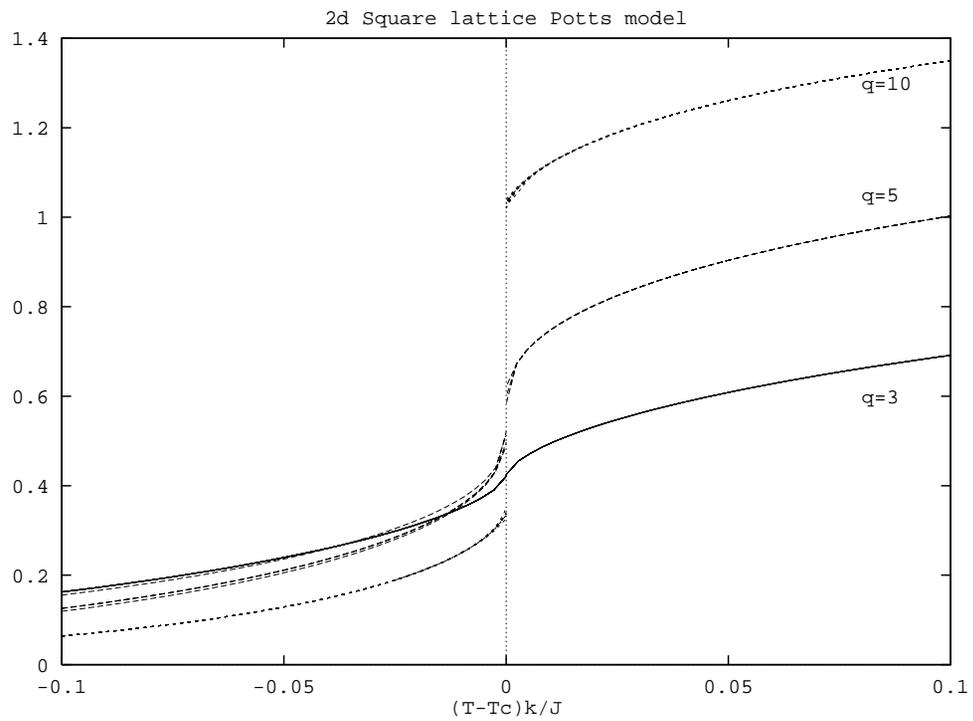

**Figure 2**

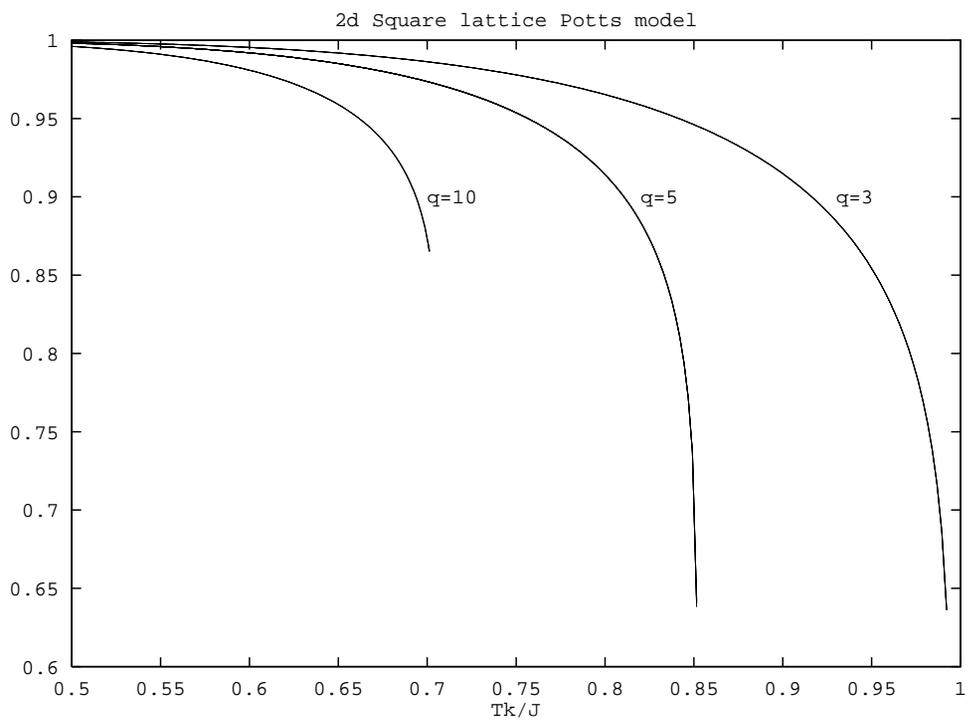

**Figure 4**